\documentstyle[12pt]{article}
\begin{document}

\begin{center}
\vglue 1.5cm
{\Large\bf Supersymmetric Photonic Signals at LEP} 
\vglue 1.5cm
{\Large Jorge L. Lopez$^1$, D.V. Nanopoulos$^{2,3}$, and A.~Zichichi$^4$}
\vglue 1cm
\begin{flushleft}
$^1$Department of Physics, Bonner Nuclear Lab, Rice University\\ 6100 Main
Street, Houston, TX 77005, USA\\
$^2$Center for Theoretical Physics, Department of Physics, Texas A\&M
University\\ College Station, TX 77843--4242, USA\\
$^3$Astroparticle Physics Group, Houston Advanced Research Center (HARC)\\
The Mitchell Campus, The Woodlands, TX 77381, USA\\
$^4$University and INFN--Bologna, Italy and CERN, 1211 Geneva 23, Switzerland\\
\end{flushleft}
\end{center}

\vglue 1cm
\begin{abstract}
We explore and contrast the single-photon and diphoton signals expected at LEP~2, that arise from neutralino-gravitino ($e^+e^-\to\chi\widetilde G\to \gamma+{\rm E_{miss}}$) and neutralino-neutralino ($e^+e^-\to\chi\chi\to\gamma\gamma+{\rm E_{miss}}$) production in supersymmetric models with a light gravitino. LEP~1
limits imply that one may observe either one, but not both, of these signals at LEP~2, depending on the values of the neutralino and gravitino masses:
single-photons for $m_\chi\raisebox{-4pt}{$\,\stackrel{\textstyle{>}}{\sim}\,$}M_Z$ and 
$m_{\widetilde G}\raisebox{-4pt}{$\,\stackrel{\textstyle{<}}{\sim}\,$}
3\times10^{-5}\,{\rm eV}$; diphotons for 
$m_\chi\raisebox{-4pt}{$\,\stackrel{\textstyle{<}}{\sim}\,$}M_Z$ 
and all allowed values of $m_{\widetilde G}$.
\end{abstract}

\vspace{1cm}
\newpage
\setcounter{page}{1}
\pagestyle{plain}
\baselineskip=14pt

Searches for supersymmetry at colliders take on a new look in the
case of models with a very light gravitino, where the lightest neutralino
($\chi^0_1\equiv\chi$) is no longer the lightest supersymmetric particle and instead decays  dominantly (in many models) into a photon and the gravitino ({\em i.e.}, $\chi\to\gamma\widetilde G$) \cite{EEN}. The $\gamma$--$\tilde\gamma$--$\widetilde G$ effective interaction is inversely proportional to the gravitino mass \cite{FayetEarly} and yields an observable inside-the-detector decay for $m_{\widetilde G}<10^3\,{\rm eV}$ \cite{Kane}. On the other hand, the gravitino mass cannot be too small, as otherwise all supersymmetric particles would be strongly produced at colliders \cite{Fayet,Dicus} or in astrophysical events \cite{astro}:
$m_{\widetilde G}>10^{-6}\,{\rm eV}$ appears required. Light gravitino scenarios were considered early on \cite{FayetEarly,EEN}, but have recently received considerable more attention because of their natural ability to explain the puzzling CDF $ee\gamma\gamma+{\rm E_{T,miss}}$ event \cite{Park} via selectron or chargino pair-production \cite{Dine,Kane,Gravitino}. Such scenarios have distinct experimental signatures, that often include one or more photons, which may be readily detected at LEP \cite{SYW,Kane,Gravitino}. 

Theoretically, light gravitinos are expected in gauge-mediated models of low-energy supersymmetry \cite{Dine}, where the gravitino mass is related to the scale of supersymmetry breaking via 
$m_{\tilde G}\approx6\times10^{-5}\,{\rm eV}\,(\Lambda_{\rm SUSY}/500
\,{\rm GeV})^2$. Special cases of gravity-mediated models may also yield light gravitinos, when the scale of local and global breaking of supersymmetry
are decoupled, as in the context of no-scale supergravity \cite{EEN,Gravitino},
in which case $m_{\tilde G}\sim (m_{1/2}/M_{Pl})^p\, M_{Pl}$, with $m_{1/2}$
the gaugino mass scale and $p\sim2$ a model-dependent constant. Our discussion
here, though, should remain largely model independent.

In the light gravitino scenario, the most accessible supersymmetric processes at LEP are $e^+e^-\to\chi\widetilde G\to \gamma+{\rm E_{miss}}$ and $e^+e^-\to\chi\chi\to\gamma\gamma+{\rm E_{miss}}$. The single-photon and diphoton processes differ in their dependence on the gravitino mass: the rate for the first process is proportional to $m^{-2}_{\widetilde G}$, whereas the second is independent of the gravitino mass. These processes also differ in their kinematical reach: $m_{\chi}<\sqrt{s}$ versus $m_{\chi}<{1\over2}\sqrt{s}$. However, one must also consider their threshold behavior, which for the single-photon process goes as $\beta^8$ \cite{Fayet} whereas for the diphoton process goes as $\beta^3$ \cite{EH}, thus compensating somewhat the different kinematical reaches. 

In this note we explore and contrast the single-photon and diphoton signals at LEP~2. The diphoton process has been considered in detail previously \cite{SYW,Kane,Gravitino}. The single-photon process was originally considered by Fayet \cite{Fayet} in the restricted case of a very light photino-like neutralino. This process was revisited in the context of LEP~1, although only
in the restricted case of a non-negligible zino component of the neutralino,
where the resonant $Z$-exchange diagram dominates \cite{DicusZ}. We have recently generalized the single-photon calculation to arbitrary center-of-mass energies and neutralino compositions, details of which appear elsewhere \cite{1gamma}. 

Let us start by considering the limits that LEP~1 imposes on the single-photon process. At $\sqrt{s}=M_Z$, this process proceeds dominantly
through $s$-channel $Z$ exchange via the coupling $Z$--$\widetilde Z$--$\widetilde G$, which is proportional to the zino component of the
neutralino $N'_{12}$.\footnote{In the notation of Ref.~\cite{HK}, the lightest neutralino can be written as $\chi\equiv\chi^0_1=N'_{11}\widetilde\gamma+N'_{12}\widetilde Z+N_{13}\widetilde H^0_1+N_{14}\widetilde H^0_2$ or alternatively as
$\chi^0_1=N_{11}\widetilde B+N_{12}\widetilde W_3+N_{13}\widetilde H^0_1+N_{14}\widetilde H^0_2$, where $N'_{11}=N_{11}\cos\theta_W+N_{12}\sin\theta_W$ and
$N'_{12}=-N_{11}\sin\theta_W+N_{12}\cos\theta_W$.} The non-resonant contributions, $s$-channel photon exchange and $t$-channel $\widetilde e_{R,L}$ exchange, are negligible unless the zino component of the neutralino is small ($N'_{12}<0.2$), in which case one must include all (resonant and non-resonant) diagrams in the calculation. The explicit expression for the cross section in the general case is given in Ref.~\cite{1gamma}. Here we limit ourselves to note its dependence on $m_{\widetilde G}$ and its threshold behavior, which is valid for all values of $\sqrt{s}$ and all neutralino compositions: 
$\sigma(e^+e^-\to\chi\widetilde G)\propto \beta^8/m^2_{\widetilde G}$, where
$\beta=(1-m^2_\chi/s)^{1/2}$. This threshold behavior results from subtle cancellations among all contributing amplitudes and was first pointed out by Fayet \cite{Fayet} in the case of pure-photino neutralinos. Dimensional analysis indicates that this cross section is of electroweak strength (or stronger) when $M^4_Z/(M^2_{\rm Pl}\,m^2_{\widetilde G})\sim1$ or $m_{\widetilde G}\sim M^2_Z/M_{\rm Pl}\sim10^{-5}\,{\rm eV}$ (or smaller).

\begin{figure}[p]
\vspace{6in}
\includegraphics{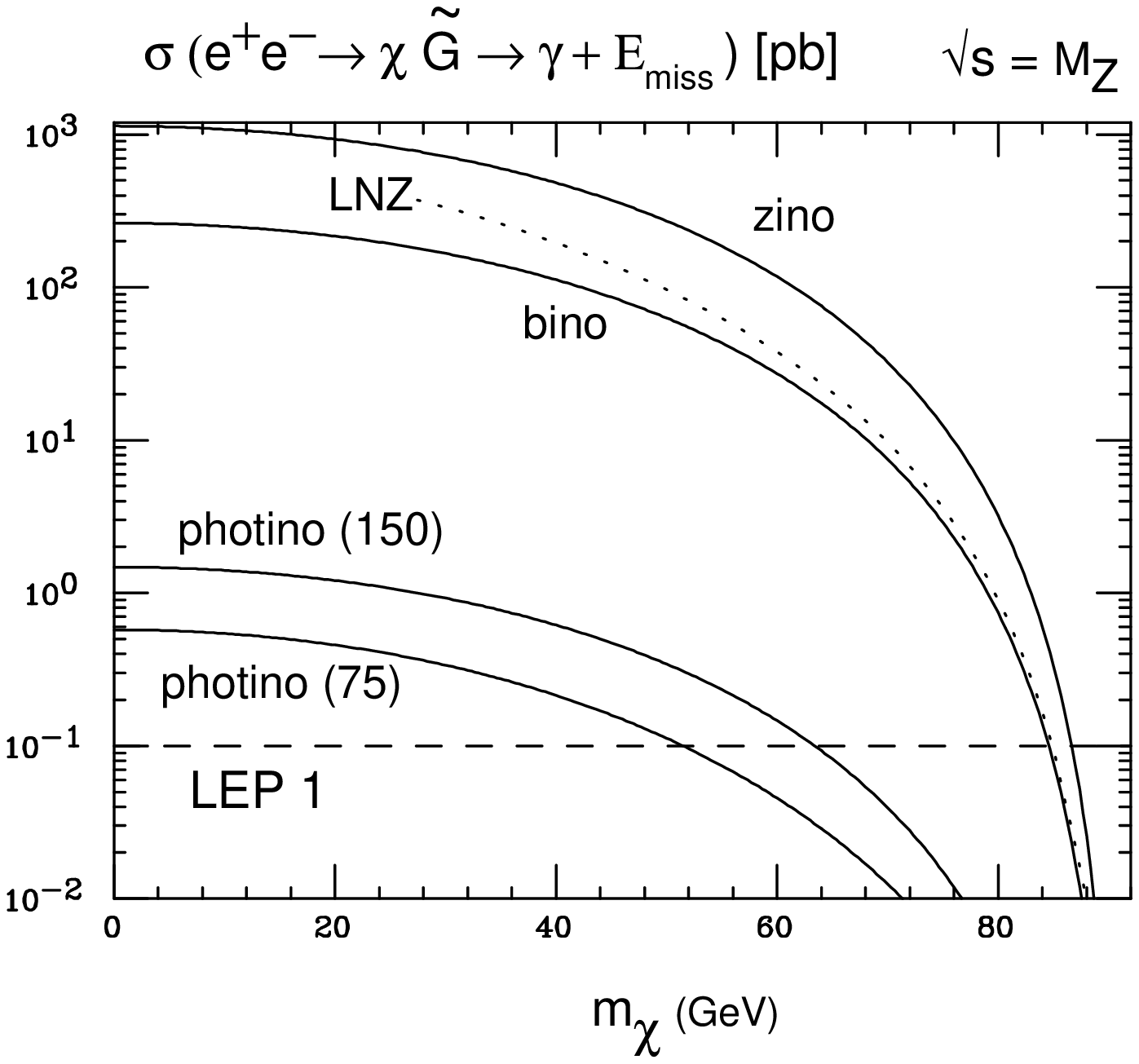}
\caption{Single-photon cross sections (in pb) from neutralino-gravitino
production at LEP~1 versus the neutralino mass ($m_\chi$) for
$m_{\widetilde G}=10^{-5}\,{\rm eV}$ and various neutralino compositions.
The `photino' curves depend on the selectron mass (75,150).
The cross sections scale like $\sigma\propto m^{-2}_{\widetilde G}$. The
dashed line represents the estimated LEP~1 upper limit.}
\label{fig:sigmaMz}
\end{figure}

A numerical evaluation of the single-photon cross section versus the neutralino mass for $m_{\widetilde G}=10^{-5}\,{\rm eV}$ is shown in Fig.~\ref{fig:sigmaMz}, for different choices of neutralino composition (`zino': $N'_{12}\approx1$; `bino': $N_{11}=1$, and `photino': $N'_{11}=1$),
and where we have assumed the typical result $B(\chi\to\gamma\widetilde G)=1$
(which assumes a (possibly small) photino admixture). In the photino case the $Z$-exchange amplitude is absent ($N'_{11}=1\Rightarrow N'_{12}=0$) and one must also specify the selectron masses which mediate the $t$-channel diagrams: we have taken the representative values $m_{\tilde e_R}=m_{\tilde e_L}=75,150\,{\rm GeV}$.  

In Fig.~\ref{fig:sigmaMz} we also show (dotted line `LNZ') the results for a well-motivated one-parameter no-scale supergravity model \cite{Gravitino,One}, which realizes the light gravitino scenario that we study here. In this model
the neutralino is mostly gaugino, but has a small higgsino component at low values of $m_\chi$, which disappears with increasing neutralino masses;
the neutralino approaches a pure bino at high neutralino masses. The selectron
masses also vary (increase) continously with the neutralino mass and are not
degenerate ({\em i.e.}, $m_{\tilde e_L}\sim 1.5\, m_{\tilde e_R}\sim2 m_\chi$).

Our particular choice of $m_{\widetilde G}=10^{-5}\,{\rm eV}$ in Fig.~\ref{fig:sigmaMz} leads to observable single-photon cross sections for $\sqrt{s}>M_Z$; otherwise the curves scale with $1/m^2_{\widetilde G}$. The dashed line indicates our estimate of the LEP~1 upper limit on the single-photon cross section of 0.1~pb \cite{LEP1}. This estimate is an amalgamation of individual experiment limits with partial LEP~1 luminosities ($\sim100\,{\rm pb}^{-1}$) and angular acceptance restrictions ($|\cos\theta_\gamma|<0.7$). Imposing our estimated LEP~1 upper limit one can obtain a lower bound on the gravitino mass as a function of the neutralino mass, which in some regions of parameter space is as strong as $m_{\widetilde G}>10^{-3}\,{\rm eV}$, but of course disappears for $m_\chi>M_Z$ \cite{1gamma}. In gauge-mediated models, such gravitino masses correspond to $\Lambda_{\rm SUSY}\sim 3\,{\rm TeV}$.

\begin{figure}[p]
\vspace{6in}
\includegraphics{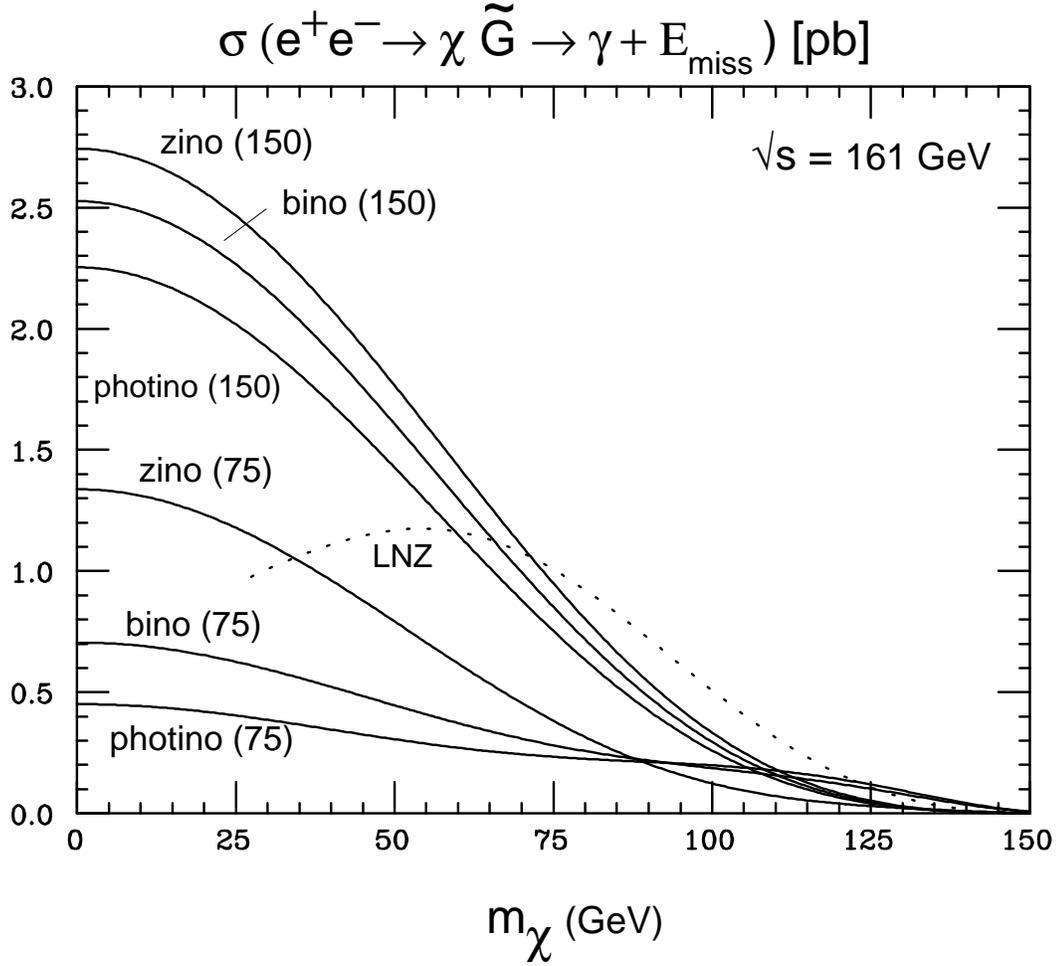}
\caption{Single-photon cross sections (in pb) from neutralino-gravitino
production at LEP~161 versus the neutralino mass ($m_\chi$) for $m_{\widetilde G}=10^{-5}\,{\rm eV}$ and various neutralino compositions.
The solid curves have a fixed value for the selectron mass (75,150), whereas the dotted curve corresponds to a one-parameter no-scale supergravity
model where the selectron masses vary continously with the neutralino mass.
The cross sections scale like $\sigma\propto m^{-2}_{\widetilde G}$.}
\label{fig:sigma161}
\end{figure}

As of this writing there are no reported excesses in the single-photon cross
sections measured at $\sqrt{s}>M_Z$. However, as it is
not clear what the actual experimental sensitivity to these processes is, we refrain from imposing further constraints from LEP~1.5 ($\sqrt{s}=(130-140)\,{\rm GeV}$) and LEP161 ($\sqrt{s}=161\,{\rm GeV}$) searches. To stimulate the experimental study of this process, in Fig.~\ref{fig:sigma161} we show the single-photon cross sections calculated at $\sqrt{s}=161\,{\rm GeV}$. Note that the cross sections increase with  increasing selectron masses (saturating at values somewhat larger than the ones shown), and conversely decrease with decreasing selectron masses. This behavior is expected: in the limit of unbroken supersymmetry ({\em i.e.}, for massless selectrons and photinos) the gravitino loses its longitudinal spin-$1\over2$ component, and therefore amplitudes involving it must vanish. This is the case in our calculations, as only the spin-$1\over2$ `goldstino' component of the gravitino becomes enhanced for light gravitino masses. Alternatively, the effective $e$--$\widetilde e$--$\widetilde G$ coupling is proportional to $m^2_{\tilde e}$ and the $t$-channel amplitude goes as $m^2_{\tilde e}/(t-m^2_{\tilde e})$, showing the dependence on $m_{\tilde e}$ and its saturation for large values of $m_{\tilde e}$; at threshold $t\to0$ and the $t$-channel amplitude becomes independent of $m_{\tilde e}$ and combines with the other amplitudes to yield the $\beta^8$ threshold behavior \cite{1gamma}.

In the case of the one-parameter model (`LNZ') a peculiar bump appears. This bump is understood in terms of the selectron masses that vary continously with the neutralino mass: at low values of $m_\chi$ the selectron masses are light and the cross section approaches the light fixed-selectron mass curves (`75'); at larger values of $m_\chi$ the selectron masses are large and the cross section approaches (and exceeds) the heavy fixed-selectron mass curves (`150'). This example brings to light some of the subtle features that
might arise in realistic models of low-energy supersymmetry.

\begin{figure}[p]
\vspace{6in}
\includegraphics{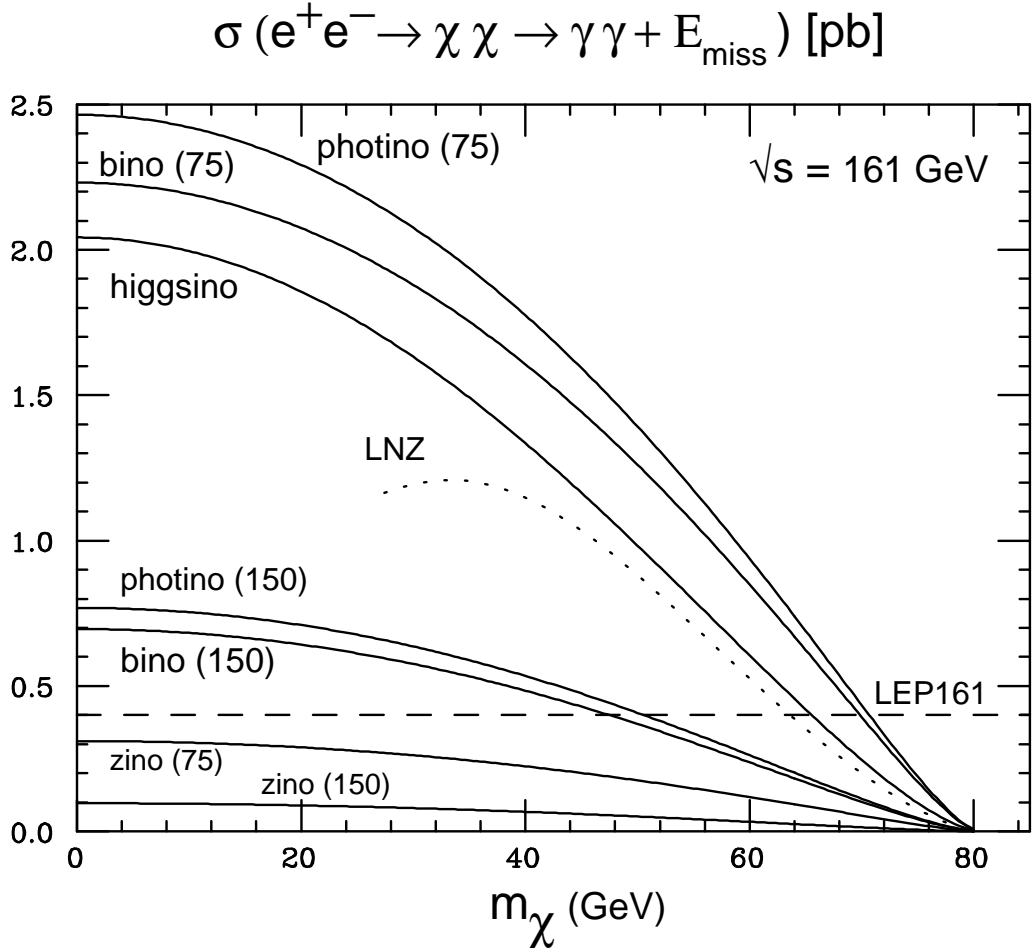}
\caption{Diphoton cross sections (in pb) from neutralino-neutralino
production at LEP161 versus the neutralino mass ($m_\chi$) for various neutralino compositions. The dependence on the selectron mass is indicated (75,150) when relevant. The dashed line represents the preliminary LEP161
upper bound.}
\label{fig:dsigma161}
\end{figure}

We now turn to the diphoton signal, which proceeds via $s$-channel $Z$-exchange
and $t$-channel selectron ($\widetilde e_{R,L}$) exchange, and does not depend
on $m_{\widetilde G}$. The $Z$-exchange contribution is present only when the neutralino has a higgsino component, whereas the $t$-channel contribution is present only when the neutralino has a gaugino component (the higgsino component couples to the electron mass). The numerical results for the diphoton cross section at $\sqrt{s}=161\,{\rm GeV}$ for various neutralino compositions are shown in Fig.~\ref{fig:dsigma161},\footnote{In Fig.~\ref{fig:dsigma161} the `higgsino' curve corresponds to the choice $N_{13}\approx1$, which maximizes the higgsino contribution. Otherwise the cross section scales as $[(N_{13})^2-(N_{14})^2]^2$.} and exhibit the expected $\beta^3$ behavior \cite{EH}. In the absence of published LEP~1 limits on the diphoton cross section (especially in the presence of substantial $\rm E_{miss}$) we turn to
higher LEP energies. Limits on acoplanar photon pairs at LEP161 have been recently released by the DELPHI, ALEPH, and OPAL Collaborations \cite{Oct8}, implying an upper bound of 0.4~pb on the diphoton cross section. Imposing this limit on the `LNZ' model entails $m_{\chi}>60\,{\rm GeV}$, with analogous limits in other regions of parameter space (see Fig.~\ref{fig:dsigma161}).

Comparing Fig.~\ref{fig:sigma161} with Fig.~\ref{fig:dsigma161}, it is amusing to note that the dependence on the selectron masses is reversed from one case to the other: the single-photon (diphoton) rate increases (decreases) with increasing selectron masses. The former behavior was explained above, the latter behavior is the usual one. The dependence on the neutralino composition is also reversed from one case to the other: zino's dominate the single-photon rate because of their $Z$-pole enhancement, bino's have some zino component and come close, while photinos have no zino component and come in last. The diphoton rate for gaugino-like neutralinos proceeds only via $t$-channel
selectron exchange and depends crucially on the coupling of left- and right-handed selectrons to neutralinos, which when examined in detail, explain 
the relative sizes of the photino, bino, and zino results in Fig.~\ref{fig:dsigma161}. 

\begin{figure}[p]
\vspace{6in}
\includegraphics{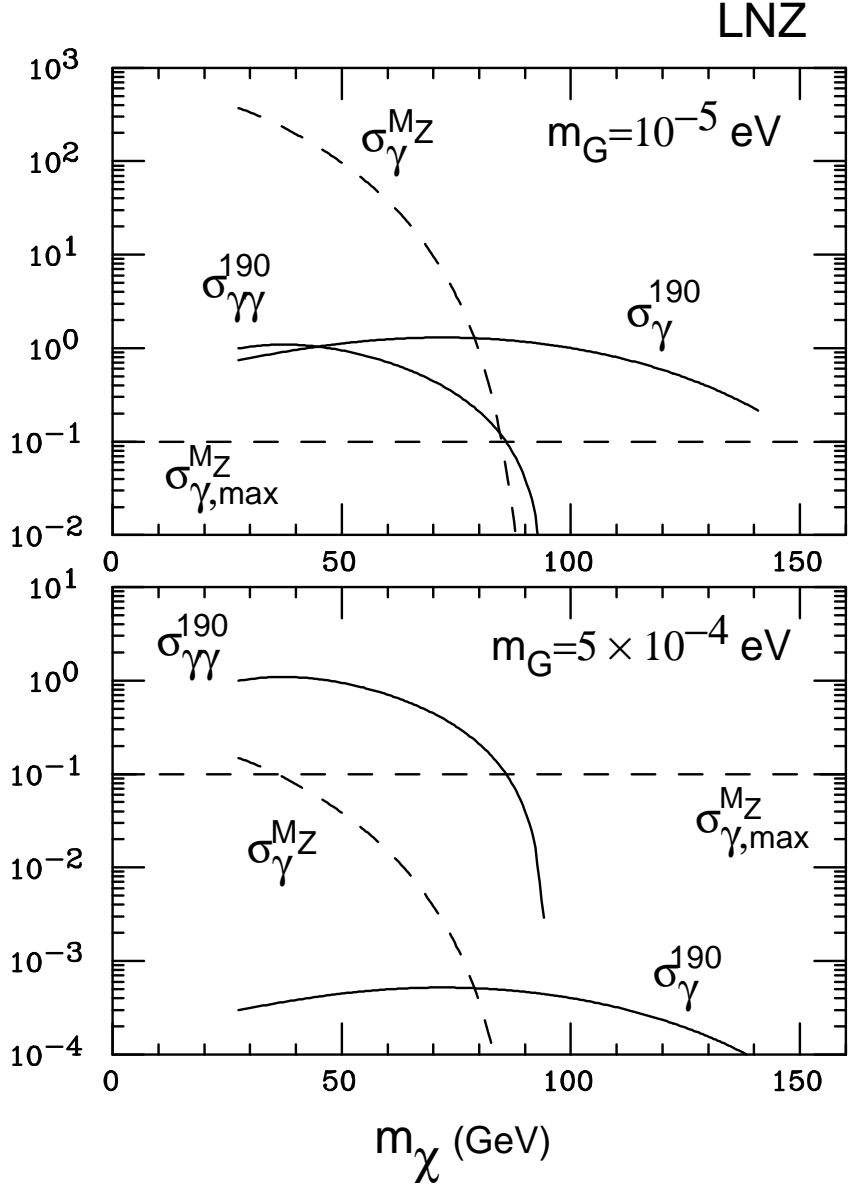}
\vspace{1cm}
\caption{Comparison of single-photon ($\sigma_\gamma^{\rm 190}$) versus diphoton ($\sigma_{\gamma\gamma}^{\rm 190}$) signals (in pb) at LEP190 as a function of the neutralino mass, for two choices of the gravitino mass. The dashed lines represent the single-photon cross section ($\sigma^{\rm M_Z}_\gamma$) and upper limit ($\sigma^{\rm M_Z}_{\gamma,{\rm max}}$) at LEP~1. The one-parameter `LNZ' model is taken here as a representative example of the two mutually exclusive scenarios that may be realized: either single photons or diphotons may be observed, but not both.}
\label{fig:1vs2LNZ}
\end{figure}

The striking point of this paper is obtained by comparing the single-photon
versus diphoton cross sections at, for example, $\sqrt{s}=190\,{\rm GeV}$,
once the LEP~1 limit on the single-photon cross section is imposed.
To exemplify the result we take as a representative example the one-parameter
(`LNZ') model \cite{Gravitino}, and plot both cross sections in Fig.~\ref{fig:1vs2LNZ}, for two values of the gravitino mass.
For $m_{\widetilde G}=10^{-5}\,{\rm eV}$ (top panel), in principle both the single-photon ($\sigma_\gamma^{\rm 190}$) and diphoton  ($\sigma_{\gamma\gamma}^{\rm 190}$) processes may be observable at LEP190. However, the LEP~1 limit on the single-photon rate ($\sigma^{\rm M_Z}_\gamma$)
can only be satisfied for $m_\chi>85\,{\rm GeV}$, and in this region the
diphoton process becomes negligible. Thus in this case one may observe only single photons. Increasing the gravitino mass to ameliorate the LEP~1 constraint on $\sigma^{\rm M_Z}_\gamma$ (to $m_{\widetilde G}=5\times10^{-4}\,{\rm eV}$, bottom panel), suppresses the single-photon rate at LEP~1 by a factor of $(50)^2$, but it suppresses the single-photon rate at LEP190 by the same factor, rendering it unobservable. However, the diphoton process at LEP190 now becomes allowed for any value of the neutralino mass (consistent with LEP~1 and LEP~1.5 limits), and this time one may observe only diphotons. Requiring a minimum observable single-photon cross section of 0.1~pb, we obtain two mutually exclusive scenarios: single-photons for $m_\chi\raisebox{-4pt}{$\,\stackrel{\textstyle{>}}{\sim}\,$}M_Z$ and 
$m_{\widetilde G}\raisebox{-4pt}{$\,\stackrel{\textstyle{<}}{\sim}\,$}
3\times10^{-5}\,{\rm eV}$; diphotons for 
$m_\chi\raisebox{-4pt}{$\,\stackrel{\textstyle{<}}{\sim}\,$}M_Z$ 
and all allowed values of $m_{\widetilde G}$.

We have verified that the same general result holds for the various other neutralino compositions that we have explored above, although in some small regions of parameter space there is a small overlap region where both single-photon and diphoton signals may be simultaneously observable. However, this may only occur for the highest LEP energies and smallest gravitino masses ($m_{\widetilde G}\sim10^{-5}\,{\rm eV}$), and only very near the diphoton
kinematical limit, where the diphoton cross section is small.

We should mention in passing that single-photon signals are also expected at the Tevatron ($p\bar p\to\chi\widetilde G$), and at even higher rates. However, large instrumental backgrounds ({\em e.g.}, $p\bar p\to W\to e\nu$, with $e$ faking a photon) may hamper such searches considerably.

In sum, we have explored the photonic signals that may be observed at LEP in
models with a light gravitino, where single-photon and diphoton signals
play a complementary role, and have the advantage over any other supersymmetric
signal of the largest reach into parameter space. 

\bigskip
\bigskip
\noindent The work of J.~L. has been supported in
part by DOE grant DE-FG05-93-ER-40717. The work of D.V.N. has been supported in
part by DOE grant DE-FG05-91-ER-40633.

\newpage

\clearpage

\clearpage

\clearpage

\clearpage

\end{document}